# Instantaneous Core Loss – Cycle-by-cycle Modeling of Power Magnetics in PWM Converters

Binyu Cui, *Student Member, IEEE*, Jun Wang, *Member, IEEE*, Xibo Yuan, *Senior Member*, IEEE, Alfonso Martinez, George Slama, Matthew Wilkowski, Ryosuke Ota, Keiji Wada

*Abstract*— Nowadays, PWM excitation is one of the most common waveforms seen by magnetic components in power electronic converters. Core loss modeling approaches, such as improved Generalized Steinmetz equation (iGSE) or the loss map based on composite waveform hypothesis (CWH), process the pulse-based excitation piecewisely, which is proven to be effective for DC/DC converters. As the additional challenge in PWM DC/AC converters, the fundamental-frequency sinewave component induces the 'major loop loss' on top of the piecewise high-frequency segments, which however cannot be modeled on a switching cycle basis by any existing methods. To address this gap, this paper proposes a novel fundamental concept, instantaneous core loss, which is the time-domain core loss observed experimentally for the first time in history. Extending the reactive voltage cancellation concept, this work presents a method to measure the instantaneous core loss, which only contains real power loss, as a function of time. Based on measurements in evaluated soft magnetic components, it was discovered that the discharging stage exhibits higher core loss than the charging stage. A modeling approach is then proposed to break down the major loop core loss, typically an average value in the literature, into the time domain to enable cycle-by-cycle modeling of core losses in PWM converters. This work enhances the fundamental understanding of the core loss process by advancing from the average model to the time-domain model.

## I. Introduction

**M**agnetic components are installed in almost all modern power electronic converters, and their power loss significantly impacts the system performance. However, there are no satisfactory first-principal models to capture all core loss mechanisms, especially under an arbitrary excitation waveform, due to the magnetic material's non-linear property and other intricate factors such as geometries. With the advancement of wide-bandgap power semiconductor devices (e.g. Silicon Carbide) and circuit topologies, a higher switching frequency appears in more high-performance power converters as a trend, which calls for better co-optimization with the passive components. Accurate loss modeling becomes increasingly important in these applications to correctly size the thermal management system, especially for the critical magnetic components, e.g. inductors and transformers. Despite the past research on core loss modeling under arbitrary waveforms, the prediction and understanding of core loss in magnetic components under PWM excitations remain challenging [1][2][3][4], especially in PWM DC/AC converters with a varying duty cycle and a fundamental component in every switching cycle. The most common modeling method used in the industry is the Steinmetz equation (SE) and its variants [5][6].

$$\overline{P_v} = k f^\alpha B^\beta \qquad (1)$$

These models are empirical equations that rely on curve fitting to acquire the Steinmetz parameters under a certain frequency range. While the original SE applies to sinusoidal waveforms, modified versions such as the improved Generalized Steinmetz equation (iGSE) extend the approach to arbitrary waveforms. However, these conventional approaches provide only average loss values after complete B-H loop closure, lacking the transient resolution needed for understanding instantaneous loss behavior during incomplete magnetic trajectories.

$$P = \frac{1}{T}\int_0^T k \left|\frac{dB}{dt}\right|^\alpha (B_{pk})^{\beta-\alpha}\, dt \qquad (2)$$

These SE equations are among the most widely applied and state-of-the-art methods for core loss modeling in industry. As shown in Fig. 1(a), the current modeling method, which is constrained by the used parameters, can only model the core loss value after the B-H trajectory completes and forms a closed B-H loop. However, under practical excitation conditions such as sinusoidal PWM (SPWM), the B–H trajectory often remains unclosed within a switching cycle, as shown in Fig. 1(b). In such cases, it becomes infeasible to evaluate the core loss dissipated along a partial trajectory segment. This limitation highlights the necessity of an instantaneous modeling approach capable of resolving core loss before the completion of a full B–H loop—for instance, estimating the loss accumulated between points O and C will become possible if the instantaneous real loss $P(t)$ can be found as a function of time.

In terms of handling partial B–H trajectories, the composite waveform hypothesis (CWH), which is the core method employed in iGSE, treats any arbitrary excitation waveform as a series of piecewise linear segments. The total core loss is then estimated by summing the individual losses associated with each segment, thereby enabling the modeling of core loss under non-sinusoidal excitations. CWH has been validated in previous works [7][8][9] providing an intuitive understanding of the core loss mechanisms, of which the basic application scenario is DC/DC converters with fixed duty cycles.

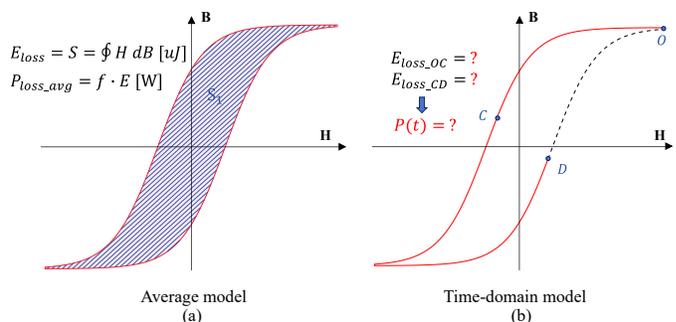

Fig. 1. Illustration of core loss modeling in B-H loops (a) state of the art (b) proposed instantaneous model $P(t)$ as a function of time

One application example of CWH in PWM converters is the loss map method proposed in [9][10][11][12] which utilizes a pre-measured look-up table that contains the core loss of a magnetic component under different operating conditions to model the overall core loss under PWM excitations. This method estimates the total core loss under PWM excitation by summing the major and minor loop losses, which correspond to the fundamental and switching frequency components, respectively. These concepts are derived from the B–H trajectory over the fundamental and switching cycles. While CWH provides a detailed framework for minor loop loss modeling, it offers limited treatment of major loop loss, which is typically approximated using an equivalent sinusoidal excitation based on the fundamental frequency component.

One of the key limitations in the existing approaches is that the major loop loss modeling, unlike minor loop loss, cannot break it down to a more granular distribution than an entire fundamental cycle. The definition of the major loop is unclear throughout different research papers. There are works defining the loops from its visual trajectory [13][14], and other papers tend to calculate the major loop loss directly from the sinusoidal excitation based on the fundamental component of the PWM excitation [11][12]. Regardless of how the major loop is defined or extracted, existing methods can only find its loss after a full cycle, once the entire B–H trajectory is closed. As a result, major loop loss is typically modeled as an average value over the cycle. However, from a physical perspective, core loss effect, including both the hysteresis and eddy current losses, occurs instantaneously along the B–H trajectory, with energy continuously converted to heat throughout, rather than a burst of heat that only occurs when the B-H loop is closed or completed. This fundamental theory indicates a conceptual disconnect in current modeling approaches, which treat core loss as a total, time-averaged quantity rather than accounting for its instantaneous variations. The limitations of average loss modeling reflect a deeper gap in the understanding of magnetic core behavior. For magnetic components with low thermal capacity, such as chip inductors and planar transformers, this instantaneous loss variations can lead to implications on the thermal cycles and component reliability [15][16]. Furthermore, emerging applications requiring high-fidelity thermal modeling and advanced control algorithms demand precise characterization of how core losses are distributed temporally within each switching cycle. This highlights the necessity for time-domain core loss modeling that can capture the instantaneous nature of energy dissipation in magnetic materials and components.

To address these limitations, this paper presents a novel direct time-domain measurement approach for instantaneous core loss characterization. Previous research has explored time-domain magnetic analysis through computational approaches such as equivalent circuit parameter extraction from instantaneous magnetizing power and harmonic analysis of power components [17][18]. Other work has focused on theoretical derivation of instant loss equations from basic magnetic principles [19]. However, these methods rely on mathematical post-processing of measured data or theoretical modeling rather than direct real-time measurement of instantaneous core loss. This work employs the **full reactive cancellation concept** [20] to enable real-time extraction of the resistive loss component without requiring post-processing procedures. This method remains applicable under incomplete B-H loops where conventional modeling approaches face limitations. The direct measurement capability enables cycle-by-cycle loss extraction during switching operations. Furthermore, this paper proposes a generic model of the instantaneous core loss that can map the core loss distribution within a cycle to the timeline, thereby solves the restraint of modeling the major loop core loss by switching cycles, which is also a supplement to the composite waveform hypothesis for DC/AC converters. The contributions of this paper are summarized as

1) The major/minor loop modeling approach under PWM excitation is revisited and verified. Through experiments, the existence of the major loop loss is verified to confirm the necessity of including it in the modeling.
2) The instantaneous core loss is defined as a fundamental concept for the first time and is empirically observed by extending the full reactive cancellation method.
3) By analyzing measured data, a generic instantaneous core loss model is proposed, which represents the time-domain loss behavior of a magnetic component. The measurement results reveal that the **discharging process is more lossy than the charging process in the tested cores**.
4) A practical workflow is proposed to extract the cycle-by-cycle major-loop loss to complete the loss map approach for practical design and modeling of magnetics in PWM converters.

II. MAJOR AND MINOR LOOP LOSS ANALYSIS

The major and minor loop loss calculation has been widely accepted to model the core loss under PWM excitation. While minor loop losses have been extensively studied and characterized in prior research [9][21], the behavior and quantification of major loop losses have not been rigorously investigated. Although major loop losses have been included in some studies, their contribution to the total core loss is often considered negligible. In the case studies where a major loop loss is, [11] reported as less than 7%, and may fall within the margin of measurement uncertainty. In many previous works, the major loop loss is approximated by the core loss under an equivalent sinusoidal excitation, with its frequency and amplitude derived via Fourier analysis. However, this approach, which is based on assumptions made in the absence of the definition of the major loop and a systematic procedure, has not been verified in practice. To address these gaps, this section will investigate the relationship between major and minor loop losses under SPWM excitation and provide further validation of the existence of major loop loss through experimental evidence. In addition, the current limitations associated with the measurement and modeling of major loop loss are also discussed.

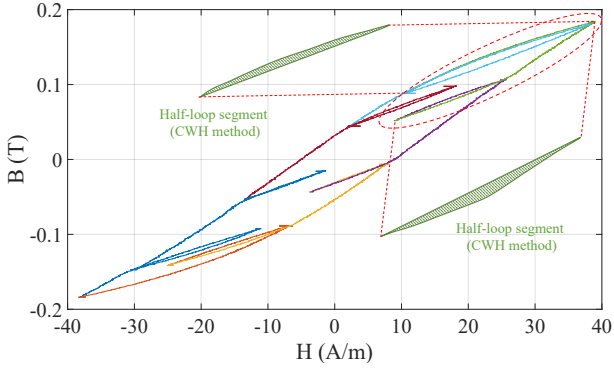

Fig. 2. SPWM B-H trajectory (modulation frequency ratio = 8) with half-loop (minor loop) segments following CWH.

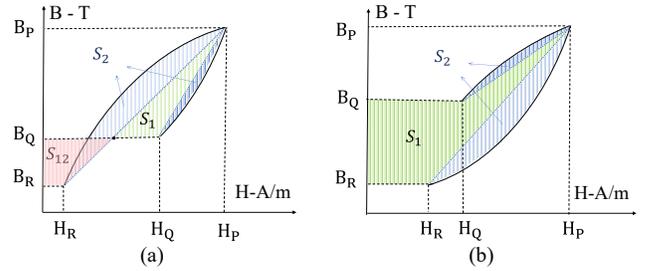

Fig. 3. B-H loop energy condition (a) Energy condition during demagnetizing (b) Energy condition during magnetizing

Fig. 2 presents an example of the B-H trajectory under the SPWM excitation. The figure shows an extreme case where the modulation frequency ratio ($f_{sw}/f_0$) is 8 to better illustrate the relationship between major and minor loops in one fundamental cycle. One of the minor B-H loops in the fundamental cycle is highlighted and used to demonstrate the practical implementation of CWH in SPWM excitation. For the minor loop loss, one of the key challenges is the unclosed B-H loop caused by the asymmetric voltage/duty cycle applied in one switching cycle as shown in the example. An unclosed loop contains the real core loss power together with the reactive power during the magnetizing and demagnetizing process. To extract the loss from the unclosed minor loop, the CWH is followed as proposed in [5] by treating the minor loop as two pieces, each of which can be seen as half of one completed B-H loop to be looked up in the loss map under the 'half loop assumption' where a half loop is associated with 50% of the loss found in the corresponding symmetric full loop. Using the loss map approach, the loss of each loop can be mapped through its magnetic flux density, magnetic field strength and frequency [11].

As it has been outlined in [21], three assumptions are followed while applying CWH.
1) The start and end points of one half-loop segment are defined by the turning point of the trajectory where the dB/dt changes its polarity.
2) For a B-H loop under symmetrical square waveform excitation, the positive half and the negative half both contribute 50% of the total energy loss.
3) 'Relaxation effect' is not considered in this situation since the common inverter topologies won't experience any constant flux density.

The **first method (Method 1)** to describe the major loop loss is as follows. By definition, the total core loss is the sum of the total minor loop loss and major loop loss. Method 1 is an indirect method for finding the total major loop loss by subtracting the minor loop loss from the total loss. As shown in Fig. 3 (a) and (b), two representative switching cycles occurring in demagnetizing and magnetizing stages in the fundamental period are presented respectively. The areas corresponding to the major and minor loop losses within each cycle are color-coded.

The blue-shaded area ($S_2$) represents the core losses associated with the B–H loop trajectory, which are known as minor loop losses that can be calculated by the CWH. The green shaded area ($S_1$) corresponds to the reactive energy stored within a switching period and the red-shaded area ($S_{12}$) represents the reactive energy being released during the demagnetization. The link between the major loop loss energy and the areas in conditions (a) and (b) is expressed as

$$E_{Q-R|(a)} = E_{major|(a)} + E_{minor|(a)} + E_{reactive|(a)}$$
$$E_{Q-R|(a)} = \int_{B_Q}^{B_P} HdB + \int_{B_P}^{B_R} HdB \quad (3)$$
$$E_{Q-R|(a)} = S_2 + S_1 - S_{12} \text{ where } E_{minor|(a)} = S_2$$
$$E_{major|(a)} + E_{reactive|(a)} = S_1 - S_{12}$$

$$E_{major|(b)} + E_{minor|(b)} + E_{reactive|(b)}$$
$$E_{R-Q|(b)} = \int_{B_R}^{B_P} HdB + \int_{B_P}^{B_Q} HdB \quad (4)$$
$$E_{major|(b)} + E_{reactive|(b)} = S_1$$

The total net energy associated with major loop loss and the reactive power of each switching cycle can be represented as the sum of the area $S_1 - S_{12}$. Since the reactive energy stored during magnetization phases is fully released during demagnetization phases upon completion of one fundamental cycle, the total major loop loss of one fundamental cycle can be expressed as:

$$\sum(S_1 - S_{12}) = E_{major\_total} + E_{reactive\_total}$$
$$\text{Since } E_{reactive\_total} = 0, \quad (5)$$
$$E_{major\_total} = \sum(S_1 - S_{12}) = E_{total\_loss} - E_{minor\_total}$$

The **second method (Method 2)** extracts the major-loop current component through FFT from a given waveform and emulates it on a standalone setup to acquire the associated equivalent major loop core loss. The detailed process of this method is illustrated through the flow chart presented in Fig. 4. As indicated, the frequency and amplitude of this equivalent sinusoidal excitation are extracted from the SPWM using the Fast Fourier Transform (FFT). After that, the experiment is carried out on a separate test set-up where the sinusoidal excitation with the matched frequency and amplitude is induced to the same component to model the major loop loss. The voltage and current measurements used the two-winding measurement to ensure only the core loss is acquired while excluding the winding loss. Through integration of the secondary winding voltage and primary winding current, the core loss of the component under test from the emulation circuit is obtained, which corresponds to the major loop loss under SPWM excitation.

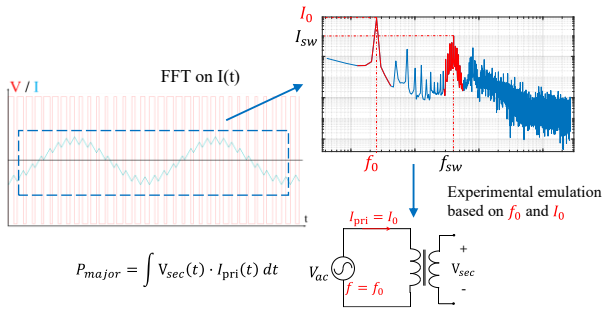

Fig. 4. Workflow of extracting major loop loss used in **Method 2**

To further validate the existence of major loop loss and to ensure that its magnitude exceeds the bounds of measurement uncertainty, a practical experiment is proposed with a practical experiment conducted under SPWM excitation on a full bridge switching configuration without any load attached. The circuit and component parameters used in the setup are detailed in TABLE I and TABLE II respectively. It should be noted that this test is conducted through the **two-winding method to** capture the core loss where the inductor under test (IUT) has two windings as indicated in TABLE II. The primary winding carries the excitation current, whereas the secondary winding is open-circuited and solely used to sense the voltage that excludes the winding loss. To enhance the detectability of the major loop loss, the experiment is deliberately configured with a reduced modulation frequency ratio $f_{sw}/f_0$, e.g. 20kHz/5kHz = 4, which serves to amplify the influence of the major loop.

Fig. 5 provides an overview of the SPWM testing platform, with all the components annotated. The voltage and current measurements in this experiment, as well as in all other tests presented in this paper, were performed using Keysight N2791A voltage probe and Keysight N2783B current probe, respectively. Fig. 6 presents representative voltage and current waveforms from the inductor under test (IUT) at a modulation ratio of 8, where the measured voltage is acquired from the secondary (sensing) winding, representing the flux density $B$ through the core and the current is measured through the primary winding.

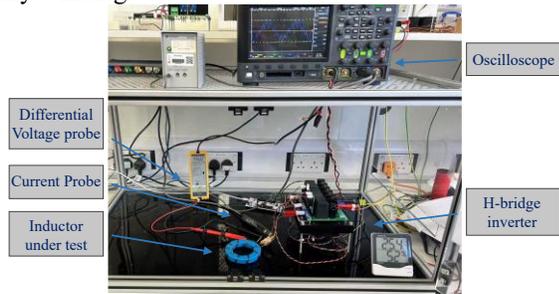

Fig. 5. H-bridge SPWM testing platform

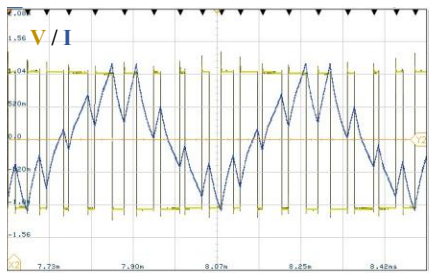

Fig. 6. Example of IUT waveforms (modulation frequency ratio = 8)

Fig. 7 illustrates the relative contributions of major and minor loop losses to the total core loss of a magnetic component under SPWM excitation. The representation of each bar is shown as follows:

1) The red bar (**Method 1**) on the left shows the total core loss ($E_{total\_loss}$) of one fundamental cycle which is calculated through the integration of the voltage and current waveform under SPWM excitation using the two-winding method.
2) The green bar (**Method 1**) on the right shows the total minor loop loss ($E_{minor\_total}$) calculated through CWH and a pre-measured database generated by the ATPT setup [22]. The difference between the red bar and the green bar (Method 1) is the total major loop loss ($E_{major\_total}$) that is calculated through method 1.
3) The blue bar represents the major loop loss that is calculated using FFT and experimental emulation, as demonstrated in **Method 2**. The difference between the height of the left bar and the cumulative height of the right bar demonstrates the difference in major loop loss calculated using **Method 1** and **Method 2** respectively.

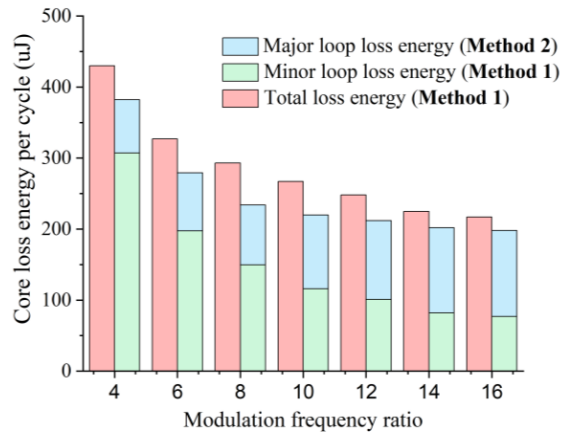

Fig. 7. Proportion Major and minor loop loss energy in one fundamental cycle proportions against different modulation frequency ratios.

The data presented in Fig. 7 serve to cross-validate two independent approaches for determining the major loop loss: (1) Subtract the independently measured total minor loop loss from the total measured loss; (2) Emulate the major loop in the PWM waveform with an equivalent sinusoidal excitation in a separate setup. Despite minor discrepancies observed at the lower modulation frequency ratio, the results obtained from both methods show good overall agreement, reinforcing the validity of the major loop loss estimation. Also, under the presented cases, the major loop loss can account for up to 60% percent of the total loss, which makes it non-negligible.

TABLE I
TEST CIRCUIT PARAMETER

| Design Parameter | Value |
| --- | --- |
| Input voltage | 20 V |
| Fundamental frequency | 2.5 kHz |
| Modulation frequency ratio | 4~16 |
| Output inductance | 264 uH |

TABLE II
PARAMETER OF INDUCTOR UNDER TEST

| Inductor parameter | Value |
|---|---|
| Core material | N87 |
| Component code | B64290L0084X087 |
| Primary winding turns | 9 |
| Secondary winding turns | 9 |

Fig. 8 and Fig. 9 present the proportions of major loop loss (unshaded area) and minor loop loss (shaded area) within the B–H loop under frequency ratios of 4 and 16 respectively. The different colored trajectory lines represent individual minor loop trajectories from different switching cycles within the fundamental period. It should be noted that under SPWM excitation, multiple minor loops are distributed throughout the B-H trajectory, contributing an additional portion of loss to the total core loss beyond what the outer boundary alone would suggest. The overlapping nature of these minor loops creates additional loss mechanisms that are not captured by conventional approaches that rely solely on the outer loop contour.

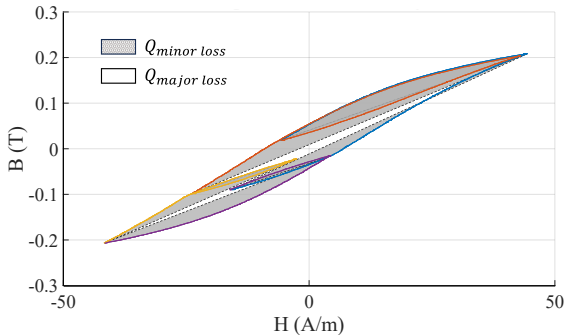
Fig. 8. Major & minor loss distribution under frequency ratio = 4.

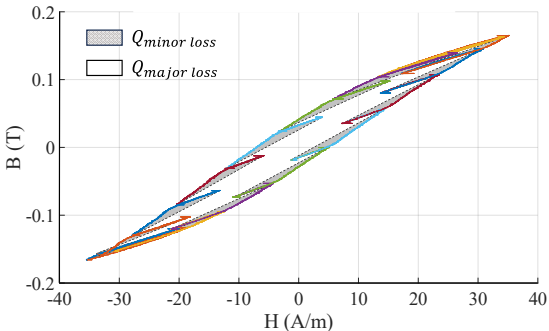
Fig. 9. Major & minor loss distribution under frequency ratio = 16.

In the extreme case with a low modulation frequency ratio of 4, the fundamental and switching frequency components are not perfectly separated in the spectrum, causing spectral leakage that shifts part of the major loop loss into minor loops, which is a limitation of the FFT-based separation. As a result, the minor loop loss constitutes a large share of the total, and a difference of over 10% arises between the calculated total core loss and the sum of the separately measured major and minor loop losses. As the modulation frequency ratio increases, the major loop loss gradually constitutes a larger fraction of the total loss while the difference rate decreases as the ratio reaches 16. While the modulation index remains constant across different modulation frequency ratios, the extracted major loop amplitude varies due to spectral leakage effects when switching harmonics are located close to the fundamental frequency at low modulation ratios. At higher frequency ratios, improved spectral separation reduces this interference, resulting in more stable major loop amplitude extraction.

As the prior work in literature, [11] shows an attempt to extract the 'instantaneous' core loss for the major loop, based on the FFT results of full cycles. The low-frequency components of the voltage and current can be regenerated, multiplied and integrated over each switching cycle to retrieve the major loop loss per cycle. However, the results of major loop loss in [11] show **negative values** in part of the switching cycles, while all the minor loop loss values are positive. This result indicates that the instantaneous major loop loss is not correctly captured through the FFT approach, since the regenerated low-frequency equivalent pair of voltage and current waveforms still involve reactive power. Additionally, due to the limitation of FFT, this approach can only be carried out at the fundamental cycle once the B-H loops are completed.

In summary, this section demonstrates (1) major loop loss exists and should not be ignored in core loss modeling, which can take up to 60% of the total loss in the evaluated cases with a high modulation frequency ratio (2) existing methods (e.g. [11]) fail to find the major loop loss on a switching cycle basis. Due to the limitations of conventional average loss modeling approaches, the major loop loss can only be accurately estimated after completing a full fundamental cycle, making it unsuitable for cycle-by-cycle loss evaluation.

III. INSTANTANEOUS CORE LOSS MEASUREMENT AND ANALYSIS

To achieve cycle-by-cycle core loss modeling, the instantaneous real power loss needs to be captured. This work proposes to extend the full reactive cancellation to capture the instantaneous core loss, which was originally developed to eliminate phase discrepancies in core loss measurements [20]. Two related techniques, partial reactive cancellation [23] and the capacitive cancellation method [24], have been reported in previous studies. However, the principles of these two methods may prevent them from capturing instantaneous core loss, in which the core loss is captured over a complete excitation cycle. In this work, the full reactive cancellation is extended to extract an in-phase voltage and current pair that directly reflects the instantaneous core loss in the time domain by excluding the reactive power contribution from the inductor. The principle is to place a reference air-core inductor ($L_{ref}$) in the circuit in series to the IUT as shown in Fig. 10.

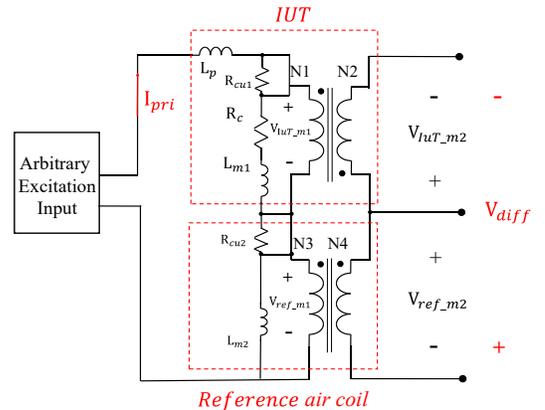
Fig. 10. Schematic of the reactive cancellation approach.

Where $L_p$ denotes the leakage inductance of the IUT; $R_{cu1}$ and $R_{cu2}$ are the winding resistances of the IUT and the reference air-core coil respectively. $R_C$ corresponds to the equivalent core loss resistance of the IUT, which is modeled as in series for analytical simplicity, similar to [20]; $L_{m1}$ and $L_{m2}$ are the magnetizing inductances of the IUT and the reference coil. A two-winding configuration is employed on both inductors, where the secondary windings are used solely for voltage sensing and carry no current. Despite the use of a transformer-like setup, no mutual inductance or inter-winding capacitance is involved. As indicated in [20], the voltage divided by the IUT and the reference coil should be identical if the full reactive cancellation is performed. If the turn ratios of N1:N2 and N3:N4 are both equal to 1, and the inductors maintain the same inductance during the measurement, the following equations stand

$$v_{IUT\_m2}(t) = R_C \cdot i_{pri}(t) + L_{m1}\frac{di_{pri}(t)}{dt} \quad (6)$$

$$v_{ref\_m2}(t) = L_{m2}\frac{di_{pri}(t)}{dt} \quad (7)$$

$$v_{diff}(t) = v_{ref_{m2}}(t) - v_{IuT_{m2}}(t) = R_C \cdot i_{pri}(t) + (L_{m2} - L_{m1})\frac{di_{pri}(t)}{dt} \quad (8)$$

If the reactive component is fully cancelled out, the voltage, $V_{diff}$, is directly related to the equivalent core loss resistor and maintains a resistive relationship with the primary current, $I_{pri}$, making it possible to directly find the instantaneous power loss when $L_{m1} = L_{m2}$

$$P(t) = i_{pri}(t)^2 \cdot R_C = v_{diff}(t) \cdot i_{pri}(t) \quad (9)$$

when $L_{m1} \neq L_{m2}$, the measurement error of

$$P_{error}(t) = (L_{m2} - L_{m1})\frac{di_{pri}(t)}{dt} \cdot i_{pri}(t) \quad (10)$$

In summary, the condition of a full reactive cancellation is $L_{m1} = L_{m2}$. When this condition is satisfied, the measured instantaneous power $P(t)$ is the real core loss converted to heat at any moment, which is a time-domain signal. The effectiveness of this method therefore critically depends on maintaining matching inductance ($L_{m1} = L_{m2}$) under actual excitation conditions. It should be noted that the equivalent circuit representation excludes winding resistance and leakage inductance from the magnetizing branch. In the two-winding measurement configuration, the secondary windings carry no current and serve solely for voltage sensing. This design isolates core losses from winding losses, as the voltage measurements exclude any resistive drops associated with copper resistance. Regarding leakage inductance, its effects can be neglected since the leakage inductance is typically much smaller than the magnetizing inductance and it's also not included in the magnetizing branch. This requirement imposes certain limitations. When the two inductances mismatch, it can lead to measurement error. In this work, the reference air coil is firstly designed to analytically match the inductance of IUT. Under small-signal conditions applied by an impedance analyzer (WK6500), the inductance difference between the IUT and the fabricated reference coil was evaluated to be 0.2% (around 0.4 μH). According to equation (10), this discrepancy results in an estimated error of approximately 8.3 mW out of 1W (0.83%) when the instantaneous power loss reaches its maximum where the current amplitude is 0.76A, as illustrated in the condition shown in Fig. 14, which is at an acceptable level. It should be noted that the inductance of the IUT varies at different operating points (e.g. different current amplitudes and frequency), due to the non-linearity of the core. To maintain the inductance matching, the number of winding turns of the reference coil is fine-tuned at each operating point to ensure that $v_{diff}(t)$ exhibits proper temporal correlation with $i_{pri}(t)$. Although adequate inductance matching is achieved for sinewave excitations in this work, especially at a relatively low frequency, it remains challenging for PWM waveforms which contain a wide spectrum of harmonic components across both the low and high frequency range. Under typical operating conditions presented in this paper, the inductance mismatch between the ferromagnetic core and reference air-core inductor remains within 1-10% throughout a complete cycle, resulting in an acceptable power error of approximately 6.8% based on equation (10). For different operating points, a few references air-core inductors with varying turn numbers or physical dimensions are fabricated to maintain the matching condition ($L_{m1} \approx L_{m2}$). It should be noted that the current work focuses on non-saturated operating regions, as saturation effects introduce additional complexities beyond the scope of this study

*A. Experimental Measurement of Instantaneous Core Loss*

The magnetic component under test is a two-winding toroidal iron powder core [25] with a turn ratio of 1:1. A reference coil with a matching inductance is connected in series with the component under test. The overview of the practical test set up for the full reactive cancellation is shown in Fig. 11. Experiments are first carried out under sinusoidal excitation to guarantee the cancellation, and a powder core, T300-26D, which the material is Mix-26 from Micrometals, is used as the IUT. The detailed specifications of the two cores are listed in TABLE III.

TABLE III
PARAMETER OF AIR CORE AND POWDER CORE

| Core Parameter | Value |
|---|---|
| Air core Se | 467.6 cm² |
| Air core le | 14.2 cm |
| Air core turns | 23:23 |
| Mix-26 core turns | 28:28 |

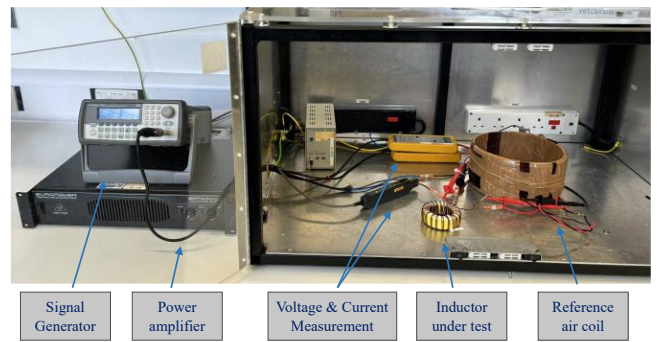

Fig. 11. Overview of the full reactive cancellation test set-up

An example of the measured waveforms is shown in Fig. 12, in which the measured $V_{diff}$ and $I_{pri}$ demonstrate proper temporal correlation, maintaining consistent polarity correspondence throughout the measurement period, indicating an effective matching of $L_{m1}$ and $L_{m2}$ and full reactive cancellation. For readability, Fig. 13 reproduces the representative oscilloscope waveforms of Fig. 12 with improved legibility, serving as a clearer demonstration of the full reactive cancellation. The effectiveness of cancellation is validated through two quantitative indicators: (a) the matching of $V_{ref}$ and $V_{IUT}$ amplitudes at current zero-crossing instants as predicted by equation (8); (b) the consistent polarity correlation between $V_{diff}$ and $I_{pri}$ throughout the measurement period indicating a resistive relationship, which ensures that the calculated instantaneous power $P(t) = V_{diff} \times I_{pri}$ remains consistently positive and resistive, thereby confirming the elimination of the reactive power component.

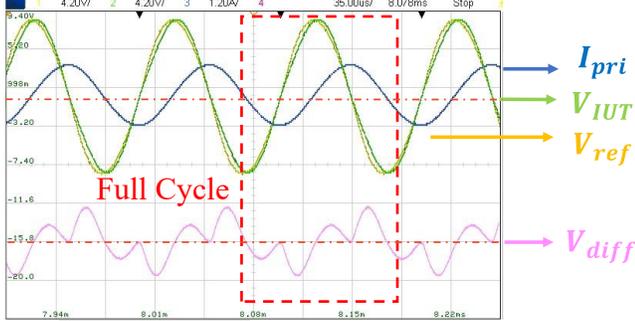

Fig. 12. Oscilloscope screenshot of Mix-26 core voltage and current waveform sinusoidal excitation.

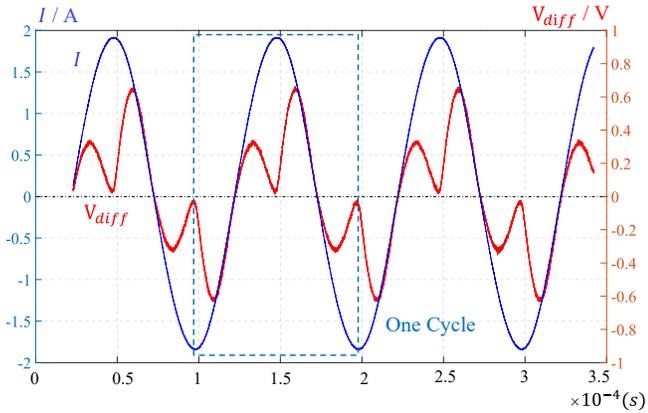

Fig. 13. Demonstration of $V_{diff}$ and current waveforms.

By integrating the $V_{diff}$ and $I_{pri}$, the instantaneous power loss of one full cycle is found through (9) and plotted in Fig. 14. As a result of full reactive cancellation, the instantaneous power loss is constantly positive, which matches the assumption of the core loss process being continuous. It can be seen that the loss dissipated in the discharging stage of the inductor is relatively higher than the charging stage. This asymmetric behavior, observed across switching cycles, reveals that core loss is not uniformly distributed within a cycle but instead varies dynamically between magnetization and demagnetization phases. It should be noted that the charging and discharging states discussed here refer to the direction of electrical energy flow based on instantaneous power, which differs from the magnetization and demagnetization processes that describe magnetic domain alignment within the core material. The charging state occurs when voltage and current have the same polarity, indicating energy storage, while the discharging state occurs when they have opposite polarities, indicating energy release.

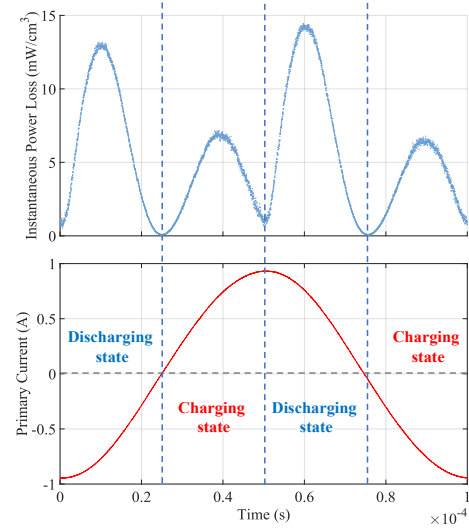

Fig. 14. Instantaneous core loss of tested Mix-26 inductor under sinusoidal excitation.

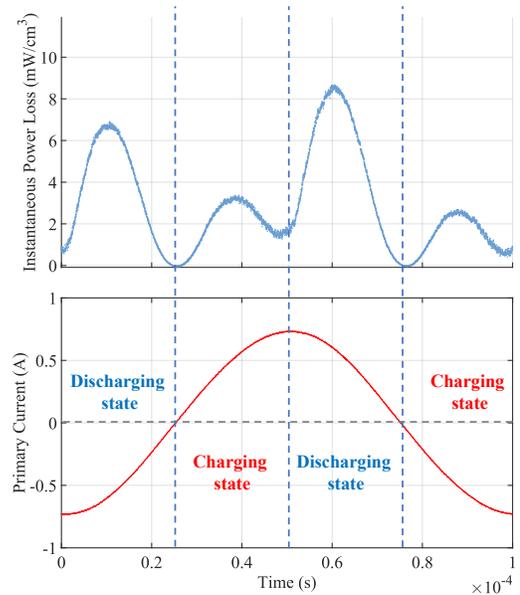

Fig. 15. Instantaneous core loss of tested N87 inductor under sinusoidal excitation.

The dynamic characteristics of the core loss is previously addressed in [26] from a physics level by attributing hysteresis loss to irreversible domain wall motion and its continuous evolution throughout the excitation process. This observed phenomenon can be interpreted as - during the discharge phase, the subset of magnetic domains that move first are those that were most resistant to alignment during the preceding charging process and therefore demanded greater energy to reorient exhibiting a high coercivity. Consequently, the discharging process witnesses a higher instantaneous core loss (as shown in Fig. 14). On the other hand, during the charging phase, the subset of magnetic domains that align first typically exhibit relatively low coercivity. These domains are more easily reoriented (i.e., they already have magnetization vectors oriented in a direction close to the applied field) and therefore require less energy for alignment. This characteristic is further confirmed on the N87 material shown in Fig. 15. The instantaneous loss shows a similar trend in the charging/discharging state.

TABLE IV
CORE LOSS ENERGY OF N87 AND MIX-26 IUT UNDER SINUSOIDAL EXCITATION

|  | N87 | Mix-26 |
|---|---|---|
| Measured total loss from regular two-winding method (from $V_{IUT\_m2}$) (uJ) | 17.7 | 36.1 |
| Measured total loss from full reactive cancellation method (from $V_{diff}$) (uJ) | 17.3 | 35.2 |
| Positive half-loop loss energy (uJ) | 8.8 | 17.8 |
| Negative half-loop loss energy (uJ) | 8.5 | 17.4 |

Fig. 16 and Fig. 17 present the instantaneous core loss distributions for Mix-26 and N87 within the B–H domain. Unlike conventional approaches, which find core loss only after completing a full fundamental cycle, this method enables the extraction of instantaneous loss at any point without closing the B–H trajectory.

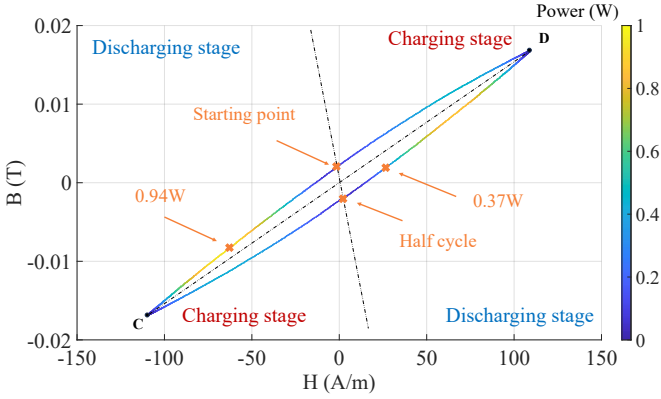

Fig. 16. Mix-26 core loss distribution in B-H domain under sinusoidal excitation.

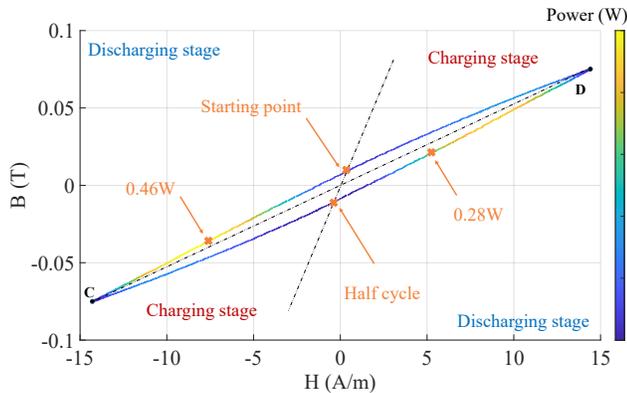

Fig. 17. N87 core loss distribution in B-H domain under sinusoidal excitation.

Moreover, the accuracy of the proposed approach is verified by comparing the total loss obtained from the inductor's measured voltage and current with the loss derived using $V_{diff}$. These results, summarized in TABLE IV, show agreement between the core loss obtained from the reactive cancellation approach and the regular two-winding method. For a full B–H trajectory, the segment from point C to point D is defined as the positive half loop, and the reverse trace from D to C is defined as the negative half loop. The results show that the energy losses associated with the positive and negative half loops are almost identical and symmetrical. This observation demonstrates the feasibility of extracting instantaneous core loss under sinusoidal excitation conditions and provides insight into the loss distribution characteristics during positive and negative half-cycles. While these results under sinusoidal excitation show relatively balanced energy distribution between half-cycles, further investigation under rectangular and PWM waveforms would be required for validation of the CWH assumption regarding equal half-loop contributions.

### B. Instantaneous core loss generic model

With the capability to measure the core loss instantaneously, more operating points are evaluated – for the Mix-26 toroidal inductor, the instantaneous core loss distributions across multiple operating points with various frequencies and voltage amplitudes are shown in Fig. 18.

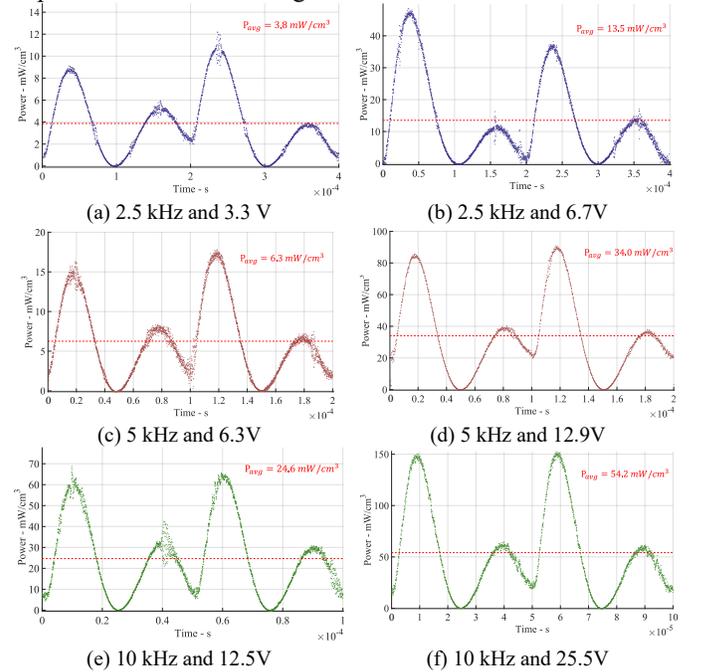

Fig. 18. Instantaneous core loss distribution of Mix-26 sample under sinusoidal excitation.

It can be seen that the time-domain instantaneous core loss waveforms show a similar pattern across various operating points. This observation leads to the generation of a generic model to represent the distribution of instantaneous loss over time of the tested component/core. Using the average power as the base, a normalized loss distribution can be derived from the acquired data points as expressed in (11). The normalized loss distribution is then averaged between the operating points into a generic model shown in Fig. 19.

$$P_{normalized} = \frac{P_{actual}}{P_{avg}} \qquad (11)$$

$$P(t) = a_0 + \sum_{n=1}^{6}[a_n \cos(nwt) + b_n \sin(nwt)] \qquad (12)$$

The distribution pattern of the generic model can be curve-fitted into a Fourier series (12) with a quality of fitting of >0.99, and the associated parameters are demonstrated in the appendix TABLE VII. This generic core loss model represents the time-domain behavior of the evaluated component. Assuming the model is consistent and representative across the majority of operating points, it can be used to break down an average loss value (e.g. energy or power retrieved from datasheets) and distribute it over time to result in the instantaneous power $P(t)$ as a function of seconds or radians. It should be noted that the generic instantaneous core loss model is developed based on the consistent pattern observed in the instantaneous loss distribution of the same material (Mix-26) under sinusoidal excitation across various operating points. For different ferrite materials or non-ferrite magnetic materials, separate generic models would need to be developed following the same methodology presented in this work. Each material exhibits unique magnetic properties, which directly influence the instantaneous loss distribution pattern.

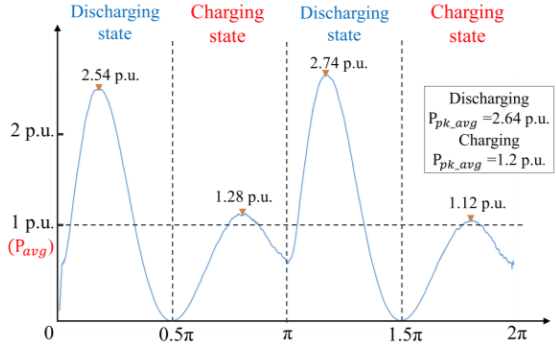

Fig. 19. Normalized instantaneous loss distribution of Mix-26 core.

IV. Cycle-by-cycle Core Loss Modeling

In a practical inductor design scenario, the proposed generic loss model can be combined with premeasured data to enable a rapid and accurate estimation of core loss on a switching cycle basis, considering both the major and minor loops. A complete workflow of estimating the core loss in a PWM inverter is illustrated in Fig. 21. As an example, an SPWM excitation described in TABLE V is applied to an inductor under test made from a T300-26D (Mix-26 material) core.

TABLE V
PWM EXCITATION PARAMETERS

| Electrical Parameter | Value |
| --- | --- |
| Input voltage | 35 V |
| Fundamental frequency | 2.5kHz |
| Modulation frequency ratio | 16 |
| Inductance of IUT | 105uH |

To find the cycle-by-cycle core loss, the low-frequency (major-loop) and high-frequency (minor-loop) components are processed in two separate flows.

- For the **low-frequency component flow**, FFT is employed to extract the fundamental-frequency sinusoidal component from the excitation voltage waveform. Using the frequency and the magnetic flux peak value of this extracted sinusoidal component, the average core loss can be retrieved from the manufacturer's datasheet [25], which yields a value of 109.3 mW/cm³. Combining the average core loss and the generic model in Fig. 19, the cycle-by-cycle low-frequency core loss distribution can be found over a fundamental cycle, as illustrated in Fig. 20 based on a total loss value of an equivalent sine wave found from the datasheet, the generic instantaneous model is applied to spread the loss into switching cycles.
- For the **high-frequency component flow**, CWH is applied to extract the positive and negative half cycles in a piecewise manner in each switching cycle. Using the premeasured loss map obtained via ATPT [22], along with the electrical loss map coordinates [21] of each half cycle, the minor loop loss for each switching cycle can be found. This pre-measured lookup table for minor loop loss can also be an artificial neural network or a database such as the open-source MagNet dataset.

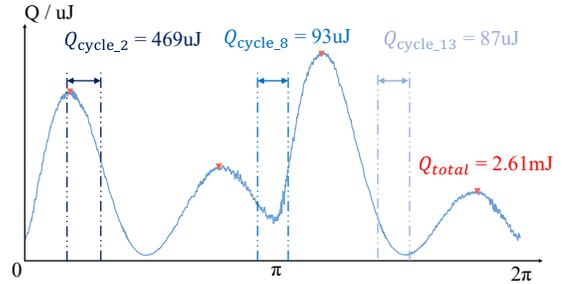

Fig. 20. Example of extracting cycle-by-cycle major loop loss from the generic model and a total loss value (e.g. $Q_{total}$ = 2.61 mJ) from the datasheet.

Through the proposed workflow, the cycle-by-cycle losses of the example case are presented in Fig. 22, which include both minor and major losses. It can be seen that, compared to the case in [11], the retrieved instantaneous loss values for the minor and major loop losses are constantly positive, indicating that only real power is captured rather than a mix of real and reactive power at a switching cycle level. TABLE VI compares major loop loss values from three independent approaches: indirect calculation (total loss subtracting minor loop loss), equivalent sinusoidal extracted from datasheet, and the proposed reactive cancellation method. The alignment among these methods validates the proposed instantaneous loss measurement and the workflow in Fig. 21. On the minor loop loss, the slight arbitrariness observed in Fig. 22 is caused by the limited number of switching cycles within a fundamental period, which is intended to amplify the effect of major loop loss in this case. In addition, the minor loop loss traces exhibit two major peaks over a fundamental cycle, which aligns with observations reported in [2], as a result of the SPWM excitation.

TABLE VI
COMPARISON OF MAJOR LOOP LOSS

| Major loop loss | Loss value (uJ) |
| --- | --- |
| Indirect measurement/calculation of whole fundamental cycle | Total loss – Minor loop loss<br>5637 - 2905 = 2732 |
| Equivalent sinusoid looked up from datasheet | 2614 |
| Equivalent sinusoid emulated in reactive cancellation circuit | 2658 |

While Fig. 21 presents the core loss decomposition into minor and major loop losses, the methodology demonstrated in this work differs from conventional approaches. Traditionally, major loop core loss is approximated as the loss induced by an equivalent low-frequency sinusoidal waveform and is typically modeled using empirical formulations such as the Steinmetz equation. This results in a single average major loss value over the entire fundamental cycle, without any temporal resolution at the switching-cycle level. In contrast, the proposed instantaneous core loss framework treats both major and minor losses as quantities that can be resolved on a switching cycle basis. This work shows that the major loop loss, although smaller in magnitude, is non-negligible and should be accounted for to achieve accurate total core loss prediction in PWM converters.

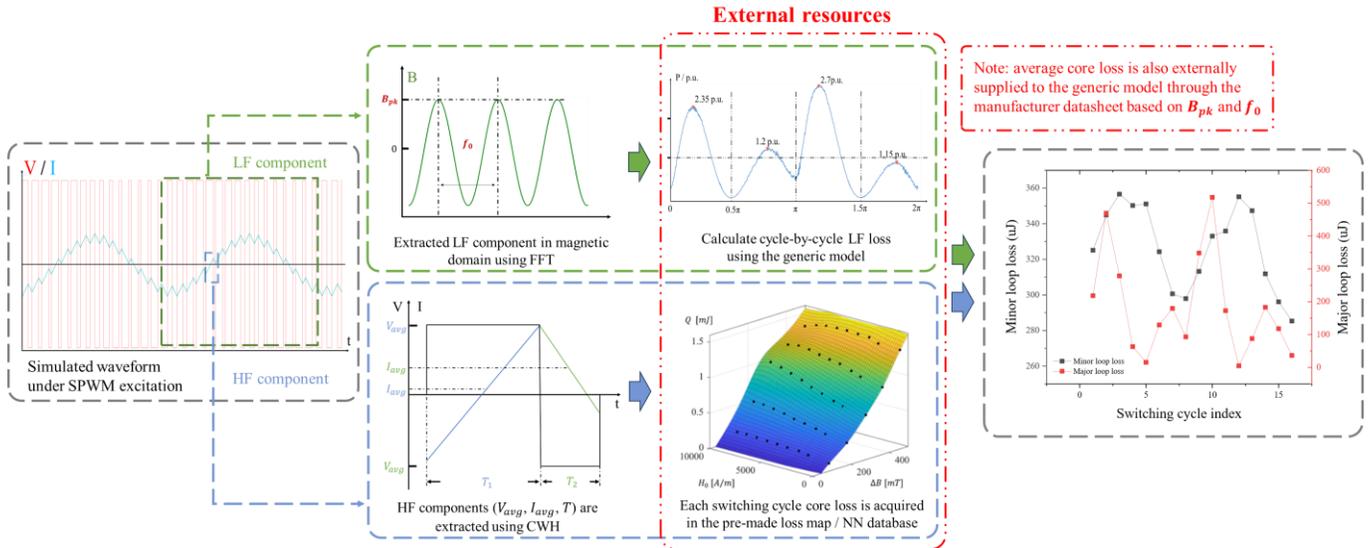

Fig. 21. Practical workflow for predicting both the major and minor loop loss on a switching cycle basis in a PWM converter

As demonstrated in Fig. 22, the major loop loss in fact arises from the accumulation of individual switching cycles throughout the fundamental cycle. This implies that major loop loss is inherently an instantaneous phenomenon, which is overlooked by the conventional method. While the average model is applicable to the situation shown in Fig. 1(a) where a full B-H loop is completed, the instantaneous model provides a viable solution for more general situations, such as the incomplete trajectories illustrated in Fig. 1(b) or the major loop loss in each switching cycle, where the average model fails to yield meaningful results.

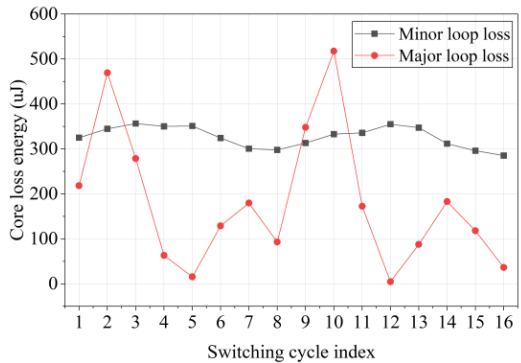

Fig. 22. Major and minor loop core loss in each switching cycle.

## V. Outlook & Summary

The concept of instantaneous core loss introduced in this work offers a novel perspective for understanding core loss dynamics in the time domain and in transient conditions. As illustrated in Fig. 23, this work bridges the conventional average model and the novel time-domain model. Specifically, the experimental extraction of major loop loss by the reactive cancellation circuit and the subsequent cycle-by-cycle modeling fills a critical gap in the CWH, where the major loop loss has historically been modeled only as an average value for the whole fundamental cycle. Additionally, measurements on representative ferrite materials reveal that discharging segments dissipate more energy than charging segments within one fundamental cycle. This implies a slight shift of the conventional anhysteretic curve in a B-H loop, typically the geometric center line, due to this asymmetry of energy loss reflected in the area during the charging/discharging stages. Accordingly, future work will be conducted to systematically assess this effect more systematically under various operating conditions.

By incorporating the instantaneous perspective, this approach is expected to enhance the data-driven and machine learning-based modeling techniques [27][28], enabling more fundamental representations of magnetic core behavior in the time domain —such as predicting $H(t)$ from $B(t)$ in transient [29]. Additionally, the proposed measurement approach provides an empirical benchmark for validating the time-domain behavior of core losses, which can be linked to the physics-based modeling and understanding [30][31]. In industrial high-power applications, this work also enables more accurate thermal cycle modeling of power magnetics, supporting improved thermal management strategies and the development of more robust component designs.

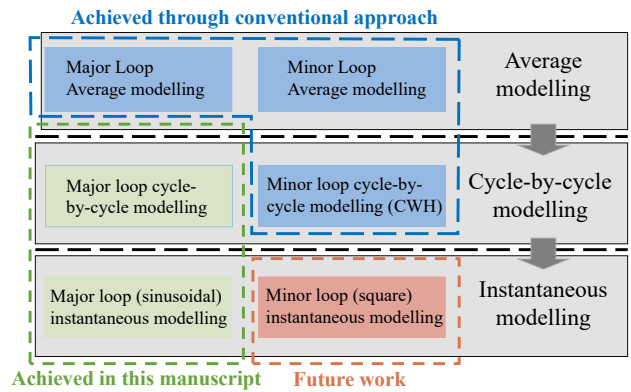

Fig. 23. Demonstration of the top-level contribution of this work.

While this work advances the state of the art, certain limitations should be acknowledged. The proposed method cannot yet achieve the extraction of instantaneous core loss under rectangular voltage. Minor loop losses are still processed as average values within each switching cycle rather than time-resolved, maintaining the established cycle-by-cycle approach using CWH, as illustrated in Fig. 23. This limitation arises from practical challenges in achieving full reactive cancellation under rectangular excitations. Due to the wide spectrum of harmonic components in rectangular voltage excitation, maintaining the inductance matching required for full reactive cancellation throughout remains challenging.

Additionally, since the concept of instantaneous core loss is introduced for the first time, this work presents the only available approach for observing it empirically. This novelty limits the ability of cross-validation with other established methods. While thermal-based techniques may offer a potential avenue for validation, existing thermal measurement approaches are unable to separate core loss from total loss. Accurate thermal measurement in fast transients would require component samples with low thermal capacity and high-precision temperature sensors, posing significant experimental challenges. Time-domain simulation tools such as ANSYS Transient Solver are available. However, these approaches are fundamentally based on empirical Steinmetz Equation and simplified assumptions that the average loss values are uniformly distributed across time steps. This solver lacks a solid factual basis and does not accurately reflect the real core loss dynamics in time domain, which differs fundamentally from the direct measurement presented in this work, making them unsuitable as validation references.

## VI. Conclusion

This paper proposes the instantaneous core loss as a fundamental concept that enables the time-domain characterization and modeling of core losses in power magnetics, moving beyond the average model. The presented full reactive cancellation setup provides a tool to capture the resistive power loss of a magnetic core in real time. Based on the time-domain loss measurements, this work reveals that there is a higher core loss exhibited at the discharging stage than at the charging stage for the tested cores. For applications in PWM converters, the existence of a major loop core loss component is validated experimentally to support the necessity of considering it. Following the measurements, generic instantaneous core loss models are extracted to represent the instantaneous behavior of magnetic components in time domain, which subsequently enables a complete cycle-by-cycle core loss estimation workflow for the power magnetics in PWM converters. This approach fills the gap of extracting the major loop loss on an instantaneous/switching cycle basis to complete the composite waveform hypothesis and the loss map approach. This work is expected to inspire and support further studies on the core behavior in transient and the fundamental physics behind it. The cycle-by-cycle core loss modeling is useful for more precise thermal design and optimization of magnetic components in power converters.

## Acknowledgements

The authors would like to thank Lukas Mueller from Micrometal (US) and Navid Rasekh from Amazon Web Services (US) for helpful comments, and Compound Semiconductor Applications Catapult (UK) for providing access to the B-H Analyzer to support this work.

## Appendix

TABLE VII
Fourier Parameters of the Fitted Function for Generic Instantaneous Core Loss Model

| | | | |
|---|---|---|---|
| $a_0$ | 0.98 | w | 0.61 |
| $a_1$ | -0.25 | $b_1$ | -0.018 |
| $a_2$ | 0.51 | $b_2$ | 0.54 |
| $a_3$ | 0.038 | $b_3$ | -0.0438 |
| $a_4$ | -0.63 | $b_4$ | 0.43 |
| $a_5$ | 0.017 | $b_5$ | -0.029 |
| $a_6$ | -0.21 | $b_6$ | 0.041 |

## References

[1] J. Muhlethaler, M. Schweizer, R. Blattmann, J. W. Kolar, and A. Ecklebe, "Optimal Design of LCL Harmonic Filters for Three-Phase PFC Rectifiers," *IEEE Trans. Power Electron.*, vol. 28, no. 7, pp. 3114–3125, July 2013.
[2] J. Wang, N. Rasekh, X. Yuan, and K. J. Dagan, "An Analytical Method for Fast Calculation of Inductor Operating Space for High-Frequency Core Loss Estimation in Two-Level and Three-Level PWM Converters," *IEEE Trans. Ind. Appl.*, vol. 57, no. 1, pp. 650–663, Jan.
[3] S. Iyasu, T. Shimizu, and K. Ishii, "A novel iron loss calculation method on power converters based on dynamic minor loop," in *2005 European Conference on Power Electronics and Applications*, Sept. 2005, p. 9 pp
[4] J. W. Kolar *et al.*, "PWM Converter Power Density Barriers," in *2007 Power Conversion Conference - Nagoya*, Apr. 2007, p. P-9-P-29.
[5] K. Venkatachalam, C. R. Sullivan, T. Abdallah, and H. Tacca, "Accurate prediction of ferrite core loss with nonsinusoidal waveforms using only Steinmetz parameters," in *2002 IEEE Workshop on Computers in Power Electronics, 2002. Proceedings.*, June 2002, pp. 36–41.
[6] J. Li, T. Abdallah, and C. R. Sullivan, "Improved calculation of core loss with nonsinusoidal waveforms," in *Conference Record of the 2001 IEEE Industry Applications Conference. 36th IAS Annual Meeting (Cat. No.01CH37248)*, Sept. 2001, pp. 2203–2210 vol.4
[7] S. Barg, K. Ammous, H. Mejbri, and A. Ammous, "An Improved Empirical Formulation for Magnetic Core Losses Estimation Under Nonsinusoidal Induction," *IEEE Trans. Power Electron.*, vol. 32, no. 3, pp. 2146–2154, Mar.
[8] T. Guillod, J. S. Lee, H. Li, S. Wang, M. Chen, and C. R. Sullivan, "Calculation of Ferrite Core Losses with Arbitrary Waveforms using the Composite Waveform Hypothesis," in *2023 IEEE Applied Power Electronics Conference and Exposition (APEC)*, Orlando, FL, USA: IEEE, Mar. 2023, pp. 1586–1593.
[9] C. R. Sullivan, J. H. Harris, and E. Herbert, "Core loss predictions for general PWM waveforms from a simplified set of measured data," in *2010 Twenty-Fifth Annual IEEE Applied Power Electronics Conference and Exposition (APEC)*, Feb. 2010, pp. 1048–1055.
[10] J. Mühlethaler, "Modeling and multi-objective optimization of inductive power components," Doctoral Thesis, ETH Zurich, 2012.
[11] H. Matsumori, T. Shimizu, K. Takano, and H. Ishii, "Evaluation of Iron Loss of AC Filter Inductor Used in Three-Phase PWM Inverters Based on an Iron Loss Analyzer," *IEEE Trans. Power Electron.*, vol. 31, no. 4, pp. 3080–3095, Apr.
[12] T. Shimizu and S. Iyasu, "A Practical Iron Loss Calculation for AC Filter Inductors Used in PWM Inverters," *IEEE Trans. Ind. Electron.*, vol. 56, no. 7, pp. 2600–2609, July 2009.
[13] K. Oda, K. Takano, and K. Wada, "Minor Loop Position and Area Measurement of Inductors for DC-DC converter Considering Excitation Process," in *2023 IEEE International Magnetic Conference - Short Papers (INTERMAG Short Papers)*, May 2023, pp. 1–2.
[14] A. Meng, J. Zhu, M. Kong, and H. He, "Modeling of Terfenol-D Biased Minor Hysteresis Loops," *IEEE Trans. Magn.*, vol. 49, no. 1, pp. 552–557, Jan.

[15] A. Lucas, R. Lebourgeois, F. Mazaleyrat, and E. Laboure, "Temperature dependence of core loss in cobalt substituted Ni–Zn–Cu ferrites," *J. Magn. Magn. Mater.*, vol. 323, no. 6, pp. 735–739, Mar.
[16] H.-S. Choi, K.-D. Kim, and J. S. Jang, "Design for reliability of ferrite for electronics materials," *Electron. Mater. Lett.*, vol. 7, no. 1, pp. 63–70, Mar.
[17] S. Divac, M. Rosić, S. Zurek, B. Koprivica, K. Chwastek, and M. Vesković, "A Methodology for Calculating the R-L Parameters of a Nonlinear Hysteretic Inductor Model in the Time Domain," *Energies*, vol. 16, no. 13, p. 5167, Jan.
[18] B. M. Koprivica and S. B. Divac, "Analysis and Modeling of Instantaneous Magnetizing Power of Ferromagnetic Cores in the Time Domain," *IEEE Magn. Lett.*, vol. 12, pp.
[19] K. Chen, "An instant loss analysis model for complex magnetic core materials," *E3S Web Conf.*, vol. 522, p.
[20] M. Mu, F. C. Lee, Q. Li, D. Gilham, and K. D. T. Ngo, "A high frequency core loss measurement method for arbitrary excitations," in *2011 Twenty-Sixth Annual IEEE Applied Power Electronics Conference and Exposition (APEC)*, Mar. 2011, pp. 157–162.
[21] J. Wang, K. J. Dagan, X. Yuan, W. Wang, and P. H. Mellor, "A Practical Approach for Core Loss Estimation of a High-Current Gapped Inductor in PWM Converters With a User-Friendly Loss Map," *IEEE Trans. Power Electron.*, vol. 34, no. 6, pp. 5697–5710, June 2019.
[22] B. Cui, J. Wang, X. Yuan, J. Aguarón, A. Martínez, and F. Cabaleiro, "Automated Triple Pulse Testbed (ATPT) 1.0 – Large-Signal Hardware-in-the-loop Characterization Platform for Power Magnetics," in *2024 IEEE 10th International Power Electronics and Motion Control Conference (IPEMC2024-ECCE Asia)*, May 2024, pp. 1613–1618.
[23] M. Mu, Q. Li, D. J. Gilham, F. C. Lee, and K. D. T. Ngo, "New Core Loss Measurement Method for High-Frequency Magnetic Materials," *IEEE Trans. Power Electron.*, vol. 29, no. 8, pp. 4374–4381, Aug.
[24] D. Hou, M. Mu, F. C. Lee, and Q. Li, "New High-Frequency Core Loss Measurement Method With Partial Cancellation Concept," *IEEE Trans. Power Electron.*, vol. 32, no. 4, pp. 2987–2994, Apr.
[25] Micrometals, "T300-26D-DataSheet.pdf," T300-26 Datasheet. [Online]. Available: https://datasheets.micrometals.com/T300-26D-DataSheet.pdf
[26] Yo Sakaki and T. Matsuoka, "Hysteresis losses in Mn-Zn ferrite cores," *IEEE Trans. Magn.*, vol. 22, no. 5, pp. 623–625, Sept.
[27] M. Chen *et al.*, "MagNet Challenge for Data-Driven Power Magnetics Modeling," *IEEE Open J. Power Electron.*, pp. 883-898, 2025.
[28] H. Li *et al.*, "MagNet: An Open-Source Database for Data-Driven Magnetic Core Loss Modeling," in *2022 IEEE Applied Power Electronics Conference and Exposition (APEC)*, Mar. 2022, pp. 588–595.
[29] H. Kwon *et al.*, "MagNetX: Extending the MagNet Database for Modeling Power Magnetics in Transient," in *2025 IEEE Applied Power Electronics Conference and Exposition (APEC)*, Mar. 2025, pp. 566–572.
[30] S. Dulal, S. B. Sohid, H. Cui, G. Gu, D. J. Costinett, and L. M. Tolbert, "A Physics-Based Circuit Model for Nonlinear Magnetic Material Characteristics," in *2024 IEEE Applied Power Electronics Conference and Exposition (APEC)*, Feb. 2024, pp. 396–401.
[31] T. Dimier and J. Biela, "Analysis of the Effect of Geometric Tolerances on Ferrite Core Losses Using a Physical Core Loss Model," in *2024 Energy Conversion Congress & Expo Europe (ECCE Europe)*, Sept. 2024, pp. 1–7.